\documentclass[10.75pt]{article}
\usepackage{geometry}
\usepackage{makeidx}
\usepackage[T1]{fontenc}
\usepackage[dvipsnames,svgnames,table]{xcolor}
\usepackage{epstopdf}
\usepackage{epsfig}
\usepackage{textcomp}
\usepackage{hyperref}
\usepackage{amsmath}
\usepackage{amssymb}
\usepackage{multirow}
\usepackage{multicol}
\usepackage{booktabs}
\usepackage{lastpage}
\usepackage{hyperref} 
\usepackage{graphicx}
\usepackage{enumerate}
\usepackage{threeparttable}
\usepackage{float}
\usepackage{array}
\usepackage{cite}
\usepackage{color,soul}
\makeatletter
\renewcommand\section{\@startsection{section}{1}{\z@}%
   {-2.5ex \@plus -1ex \@minus -.2ex}%
   {2.3ex \@plus.2ex}%
   {\normalfont\large\bfseries}}
\renewcommand\subsection{\@startsection{subsection}{1}{\z@}%
   {-2.5ex \@plus -1ex \@minus -.2ex}%
   {2.3ex \@plus.2ex}%
   {\small\bfseries}}
\begin{document}
\title{\vskip -2.5 cm\textbf{Unveiling triangular correlation of angular deviation in muon scattering tomography by means of GEANT4 simulations}}
\medskip
\author{\small A. Ilker Topuz$^{1,2}$, Madis Kiisk$^{1,3}$, Andrea Giammanco$^{2}$, and Mart Magi$^{3}$}
\medskip
\date{\small$^1$Institute of Physics, University of Tartu, W. Ostwaldi 1, 50411, Tartu, Estonia\\
$^2$Centre for Cosmology, Particle Physics and Phenomenology, Universit\'e catholique de Louvain, Chemin du Cyclotron 2, B-1348 Louvain-la-Neuve, Belgium\\
$^3$GScan OU, Maealuse 2/1, 12618 Tallinn, Estonia}
\maketitle
\begin{abstract}
The angular deviation commonly represented by the scattering angle generally serves to provide the characteristic discrimination in the muon scattering tomography. The regular procedure to determine the scattering angle comprises the collection of exactly four hit locations in four detector layers among which two top detector layers are utilized to construct the first vector, whereas the second vector is built by using two bottom detector layers. Although this procedure acts to classify the target volumes in the tomographic systems based on the muon scattering, the scattering angle obtained through the usual methodology founded on four detector layers is dubious for not yielding any information about the position of target volume. Nonetheless, the same set of four detector layers also imparts the possibility of splitting the scattering angle into two separate angles by creating a triangular correlation in such a way that the scattering angle is referred to an exterior angle, whereas the separate angles are considered the interior opposite angles that are not neighboring this exterior angle. In this study, we first show that a combination of three detector layers out of four fulfills the calculation of the interior opposite angles. Then, by employing the GEANT4 simulations over our tomographic configuration composed of three plastic scintillators in either section, we demonstrate that the interior opposite angles differ towards the vertical spatial variation, while the exterior angle approximately remains constant, thereby implying a beneficial feature to be used for the image reconstruction purposes.
\end{abstract}
\section{Introduction}
In the muon scattering tomography~\cite{pesente2009first, Checchia_2016, procureur2018muon, bonechi2020atmospheric}, the scattering angle due to the volume-of-interest (VOI) and its associated statistics act as the principal variables in order to discriminate as well as to reconstruct the corresponding VOIs in the image reconstruction techniques such as Point-of-Closest Approach (POCA)~\cite{schultz2003cosmic, bandieramonte2013automated, yu2013preliminary, liu2018muon, yang2019novel,zeng2020principle, liu2021muon}.  As specified by the conventional tomographic configurations based on the muon scattering~\cite{borozdin2003radiographic}, the entire detection system regularly includes a bottom hodoscope below the VOI in addition to a top hodoscope above the VOI on the condition of multiple detector layers present at each hodoscope~\cite{bandieramonte2013automated, yu2013preliminary, zeng2020principle}. In these tomographic setups hinged on the muon scattering, the scattering angle is commonly computed by constructing a vector~\cite{carlisle2012multiple, nugent2017multiple, poulson2019application} founded on two hit locations at two distinct detector layers within every hodoscope. Although this prevalent procedure that is devoted to calculate the characteristic angular deviation anywise serves to differentiate the corresponding VOIs or to produce their radiographic images, the angular variation towards the spatial change of the same VOIs is not definitely assured by using the habitual definition of the scattering angle, which also means that the regular utilization of the two hit locations at each section might not yield any further information towards the position change of the target material.

In the present study, motivated by this question mark about the angular alteration by means of the common definition versus the spatial variation, we first show that the same set of four hit locations collected from the two detector layers at every hodoscope might lead to split the scattering angle into two opposite angles by forming a triangular correlation where the scattering angle is considered an exterior angle, while the two separate angles by definition are interior angles that are not neighboring the scattering angle. In the second place, we perform a series of GEANT4 simulations~\cite{agostinelli2003geant4} by changing the vertical position of the VOI made out of stainless steel within our tomographic scheme~\cite{georgadze2021method} consisting of three plastic scintillators manufactured of polyvinyl toluene and we demonstrate that the interior opposite angles vary depending on the VOI location, whereas the scattering angle that is expressed according to the regular definition does not yield a significant difference despite this spatial change. Last but not least, the triangular correlation between the scattering angle and the interior opposite angles is corroborated by the equality between the scattering angle and the sum of these non-adjacent angles via our GEANT simulations. The current study is organized as follows. In section~\ref{triangular correlation}, we define the scattering angle as well as the interior separate angles in accordance with the triangular correlation by delineating over our tomographic configuration, and section~\ref{simulation scheme} is composed of our simulation schemes in order to explore the position sensitivity of the scattering angle as well as the opposite interior angles obtained by splitting the scattering angle. While we exhibit our simulation results in section~\ref{simulation outcomes}, we draw our conclusions in section~\ref{conclusion}.
\section{Triangular correlation}
\begin{figure}[H]
\begin{center}
\includegraphics[width=7.5cm]{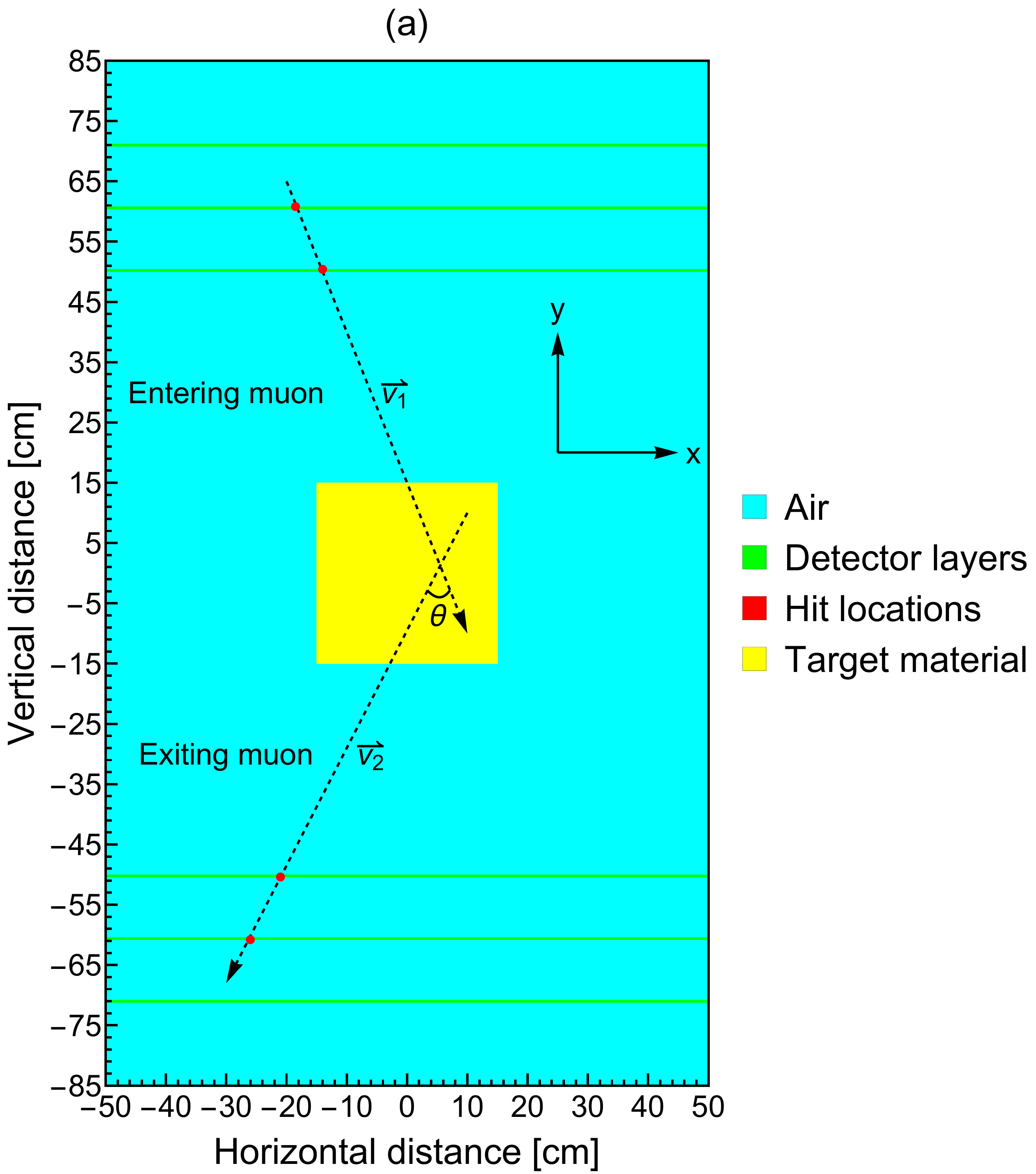}
\hskip 1cm
\includegraphics[width=7.5cm]{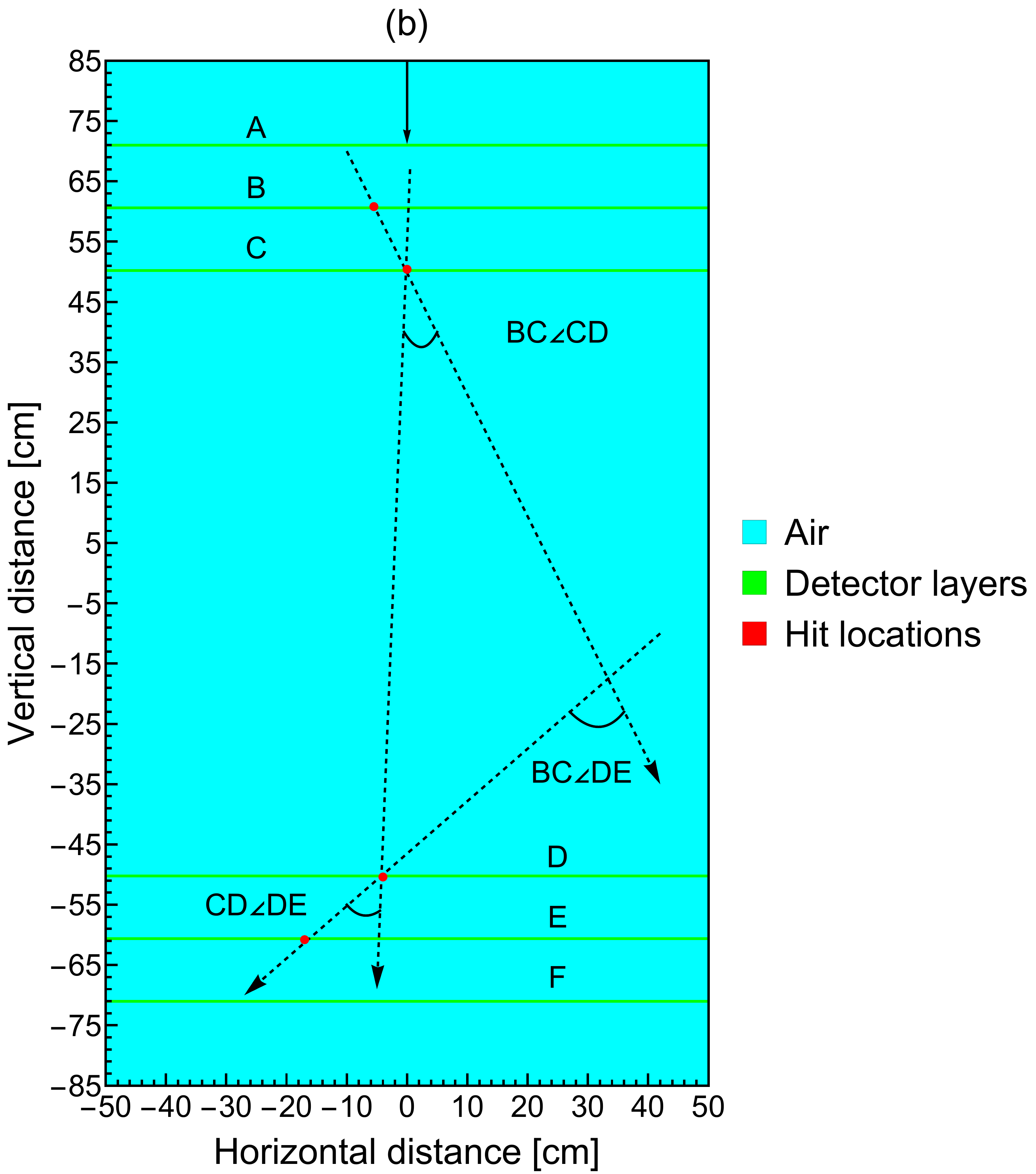}
\caption{Delineation of angular deviation due to the target volume in our tomographic scheme: (a) scattering angle denoted by $\theta$ and (b) triangular correlation between $\theta=\rm BC\angle DE$ and the interior angles denoted by $\rm BC\angle CD$ and $\rm CD\angle DE$ after splitting.}
\label{angular}
\end{center}
\end{figure}

To begin with, our tomographic setup is depicted in Fig.~\ref{angular} (a) where the scattering angle indicated by $\theta$ is determined by building a vector at each section, the components of which are obtained through the hit locations on two detector layers. The scattering angle might be split into two opposite angles by setting up a triangular correlation as illustrated Fig.~\ref{angular} (b) where the exterior angle referred to the scattering angle is equal to the superposition of the two non-adjacent angles.
\label{triangular correlation}
\vskip -0.25cm
By reminding that the capital letters listed as $\rm A, B, C, D, E,$ and $\rm F$ in Fig.~\ref{angular} (b) point to the hit locations in the specific detector layers, the conventional scattering angle denoted by $\theta$ that also refers to the exterior angle is commonly defined as written in~\cite{carlisle2012multiple, nugent2017multiple, poulson2019application}
\begin{equation}
\theta=\rm BC\angle DE=\rm BC\angle CD + \rm CD\angle DE=\arccos\left(\frac{\overrightarrow{\rm BC}\cdot\overrightarrow{\rm DE}}{\left|BC\right|\left|DE\right|}\right)
\label{superpose}
\end{equation}
The same set of four hit locations also gives access to compute two opposite interior angles as expressed in 
\begin{equation}
\rm BC\angle CD=\arccos\left(\frac{\overrightarrow{\rm  BC}\cdot\overrightarrow{\rm  CD}}{\left|BC\right|\left|CD\right|}\right)
\end{equation}
and
\begin{equation}
\rm CD\angle DE=\arccos\left(\frac{\overrightarrow{\rm  CD}\cdot\overrightarrow{\rm  DE}}{\left|CD\right|\left|DE\right|}\right)
\end{equation}
It is worth mentioning that the computation of the interior angles indicated by $\rm  BC\angle CD$ and $\rm  CD\angle DE$ does not require any further data collection from the detector layers since the same set of four hit locations are already mandatory to calculate the scattering angle, and three hit points out of four are sufficient in order to determine these non-adjacent angles. The average angular deviation of any combination, i.e  $\overline{x\angle y}$, at a given energy value is determined by averaging over $N$ number of the non-absorbed/non-decayed muons as defined in 
\begin{equation}
\overline{x\angle y}=\frac{1}{N}\sum^{N}_{i=1}(x\angle y)_{i}
\end{equation}
\section{Simulation scheme for position sensitivity}
\label{simulation scheme}
\begin{figure}[H]
\begin{center}
\includegraphics[width=5cm]{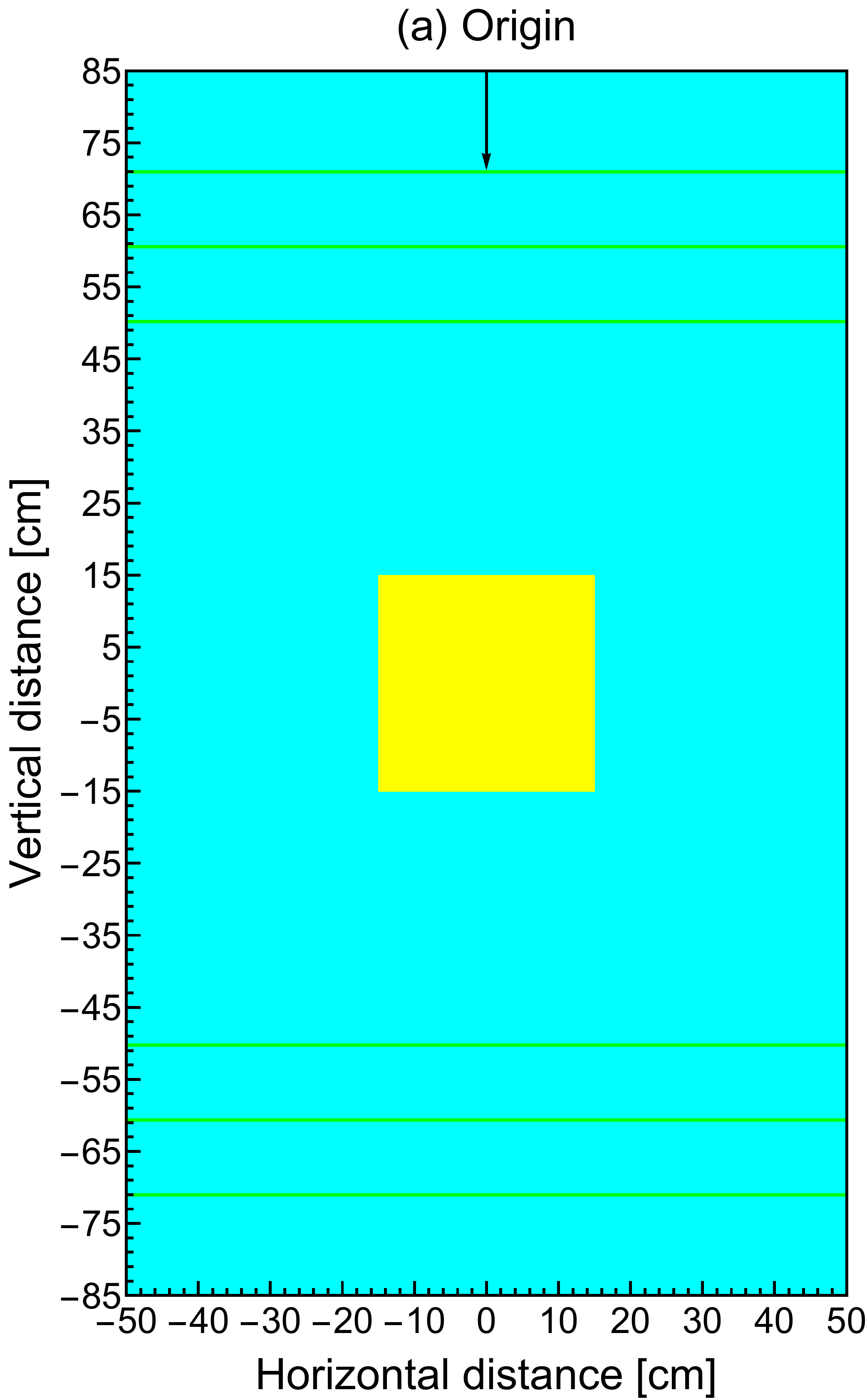}
\hskip 0.5cm
\includegraphics[width=5cm]{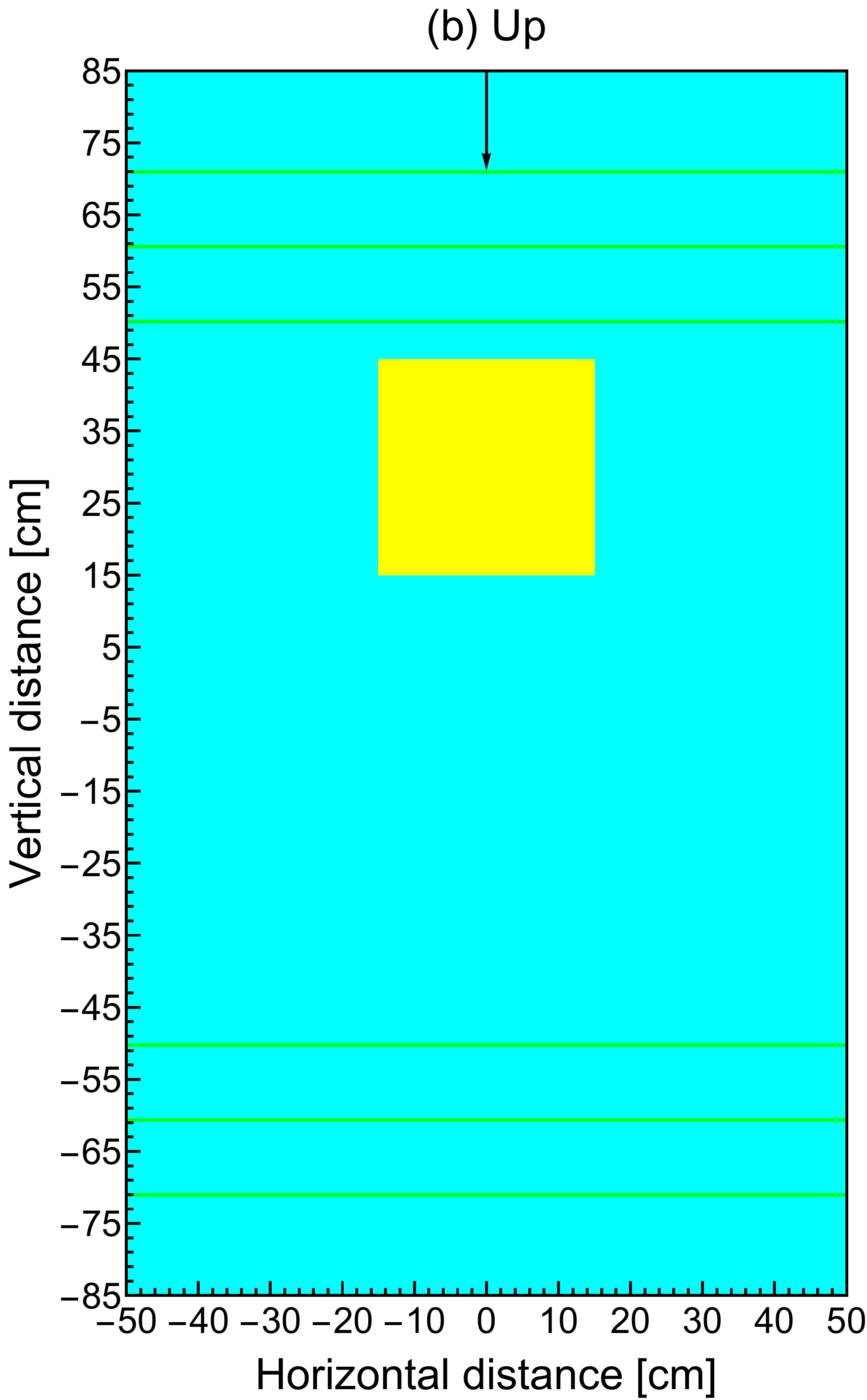}
\hskip 0.5cm
\includegraphics[width=5cm]{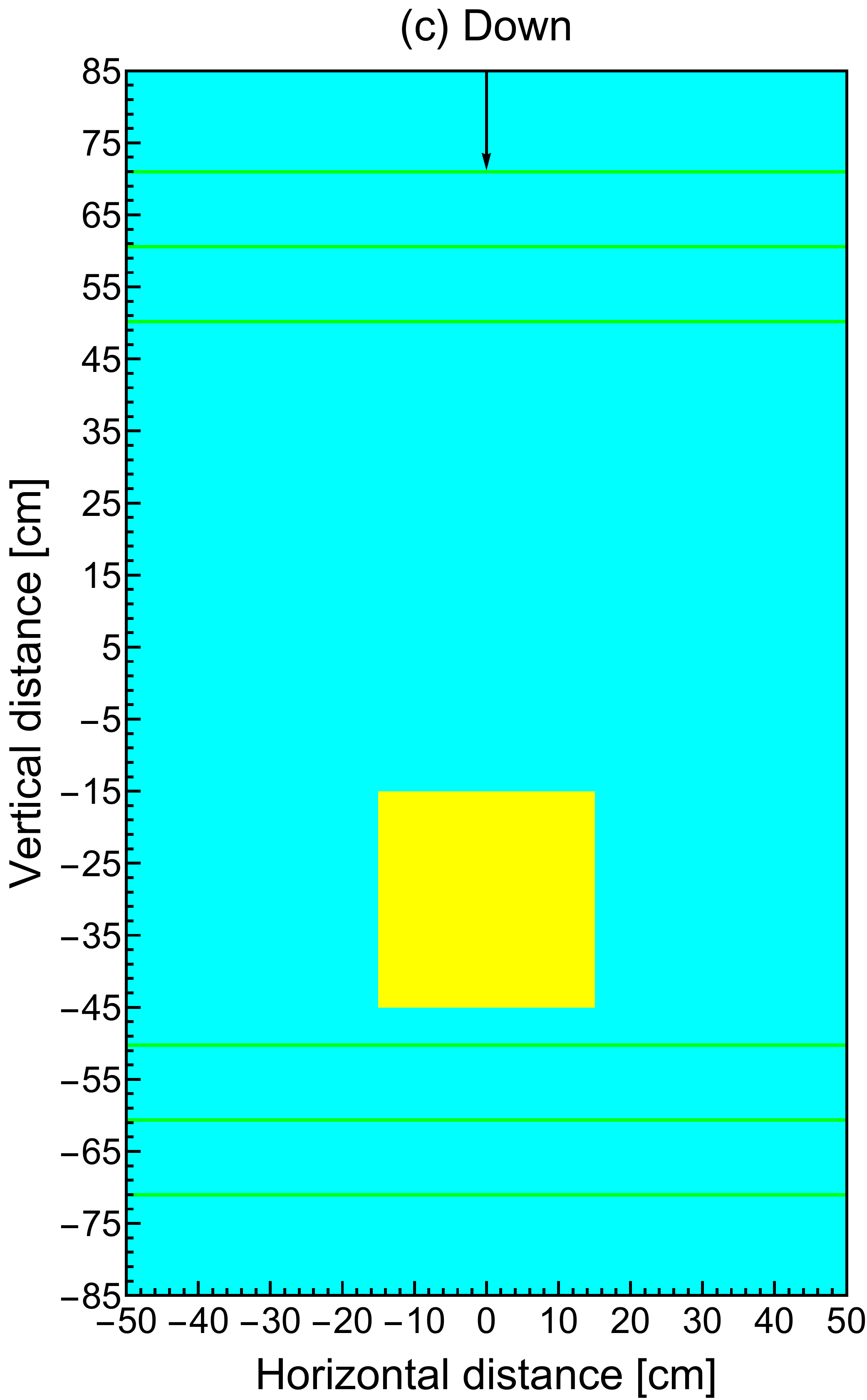}
\caption{Simulation schemes for the position sensitivity by using three different vertical VOI centers with (a) origin at (0, 0), (b) up at (30, 0), and (c) down at (-30, 0) in cm.}
\label{position}
\end{center}
\end{figure}

Following the definition of the triangular correlation and the associated angles of this correlation collected based on the tracked hits from the detector layers, we perform a sequence of GEANT4 simulations in order to verify the triangular correlation as well as to testify for the position sensitivity. We define three position cases in cm that consist of origin, up, and down as delineated in Fig.~\ref{position} (a)-(c) where (a) shows the case called origin and the center of the VOI is located at (0, 0), (b) demonstrates the case labeled up and the center of VOI is moved to (30, 0), and (c) depicts the case termed down and the center of VOI is situated at (-30, 0). Apart from the VOI position, the VOI material is stainless steel with a cubic volume of $30\times30\times30$ $\rm cm^{3}$.
\vskip -0.25cm
To concisely summarize, our tomographic setup in GEANT4 simulations is composed of three plastic scintillators made out of polyvinyl toluene with the dimensions of $100\times0.4\times100$ $\rm cm^{3}$ at every section. We utilize a central mono-directional uniform muon beam as indicated by a downward black arrow in Fig.~\ref{position} (a)-(c), and the uniform energy distribution~\cite{anghel2015plastic} lies on an interval between 0.1 and 8 GeV for the reason of more favorable numerical stability. The number of the simulated muons in each defined position is $10^{5}$. The tomographic components in the GEANT4 simulations are defined in agreement with the G4/NIST database, and the preferred physics list is FTFP$\_$BERT. The simulation features are listed in Table~\ref{features}.
\begin{table}[H]
\begin{center}
\begin{footnotesize}
\caption{Simulation features.}
\begin{tabular}{c c}
\toprule
\toprule
Particle & $\mbox{\textmu}^{-}$\\
Beam direction & Vertical\\
Momentum direction & (0, -1, 0)\\
Source geometry & Planar\\
Initial position (cm) & ([-0.5, 0.5], 85, [-0.5, 0.5])\\
Number of particles & $10^{5}$\\
Energy distribution & Uniform\\
Energy interval (GeV) & [0, 8]\\
Bin step length (GeV) & 0.5\\
Energy cut-off (GeV) & 0.1\\
Target material & Stainless steel\\
Target geometry & Cube\\
Target size (cm) & 30\\ 
Material database & G4/NIST\\
Reference physics list & FTFP$\_$BERT\\
\bottomrule
\bottomrule
\label{features}
\end{tabular}
\end{footnotesize}
\end{center}
\end{table}
\vskip -0.25cm
The muon tracking is accomplished by G4Step, and the tracked hit locations are post-processed by the support of a Python script where the scattering angle and the interior non-adjacent angles are initially computed for every single non-absorbed/non-decayed muon, then the uniform energy spectrum limited by 0.1 and 8 GeV is divided into 16 bins by marching with a step of 0.5 GeV, and each obtained energy bin is labeled with the central point in the energy sub-interval. Finally, the determined angles are averaged for the associated energy bins.
\section{Simulation outcomes}
\begin{figure}[H]
\begin{center}
\includegraphics[width=7.34cm]{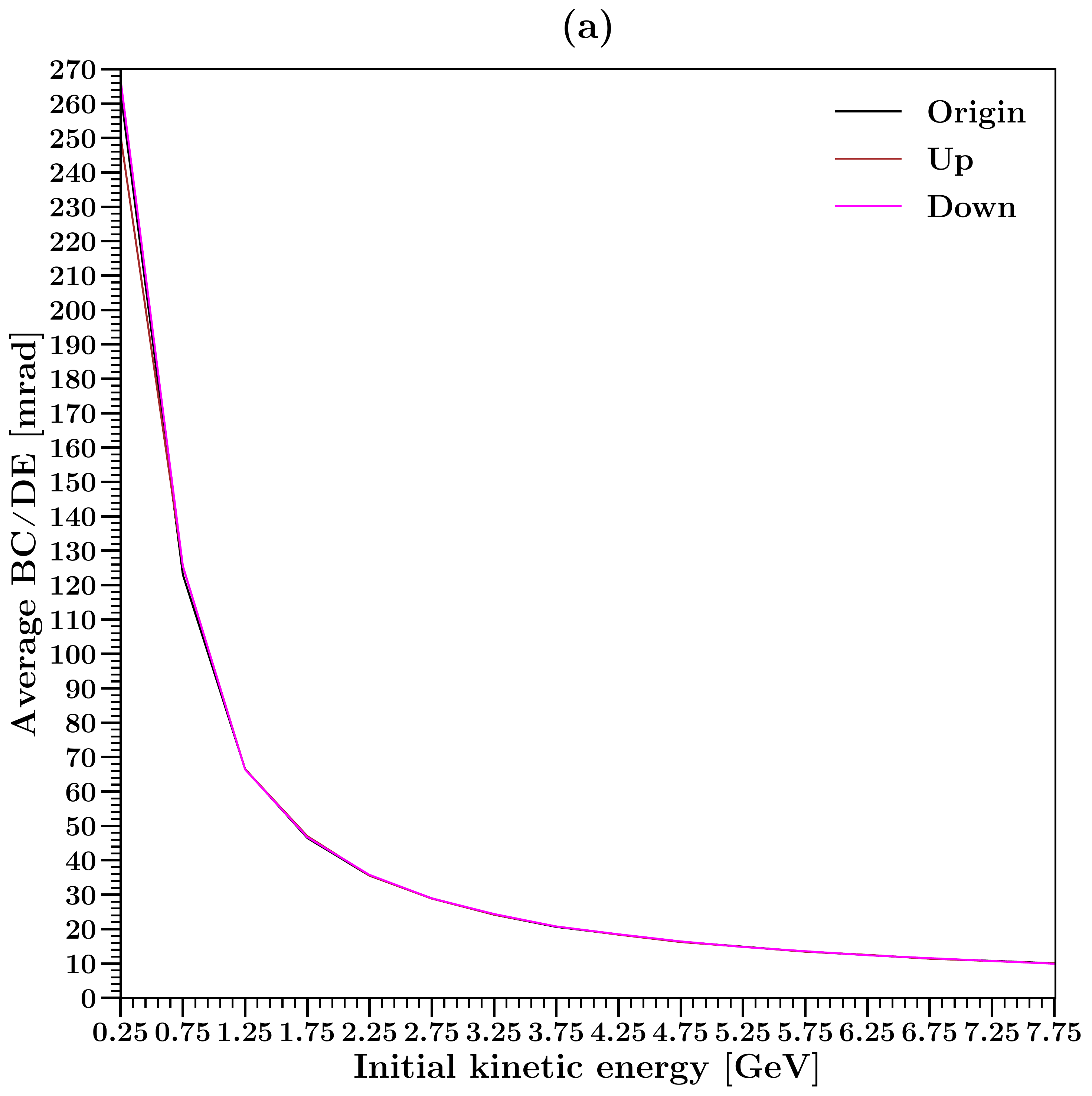}
\hskip 0.3cm
\includegraphics[width=7.34cm]{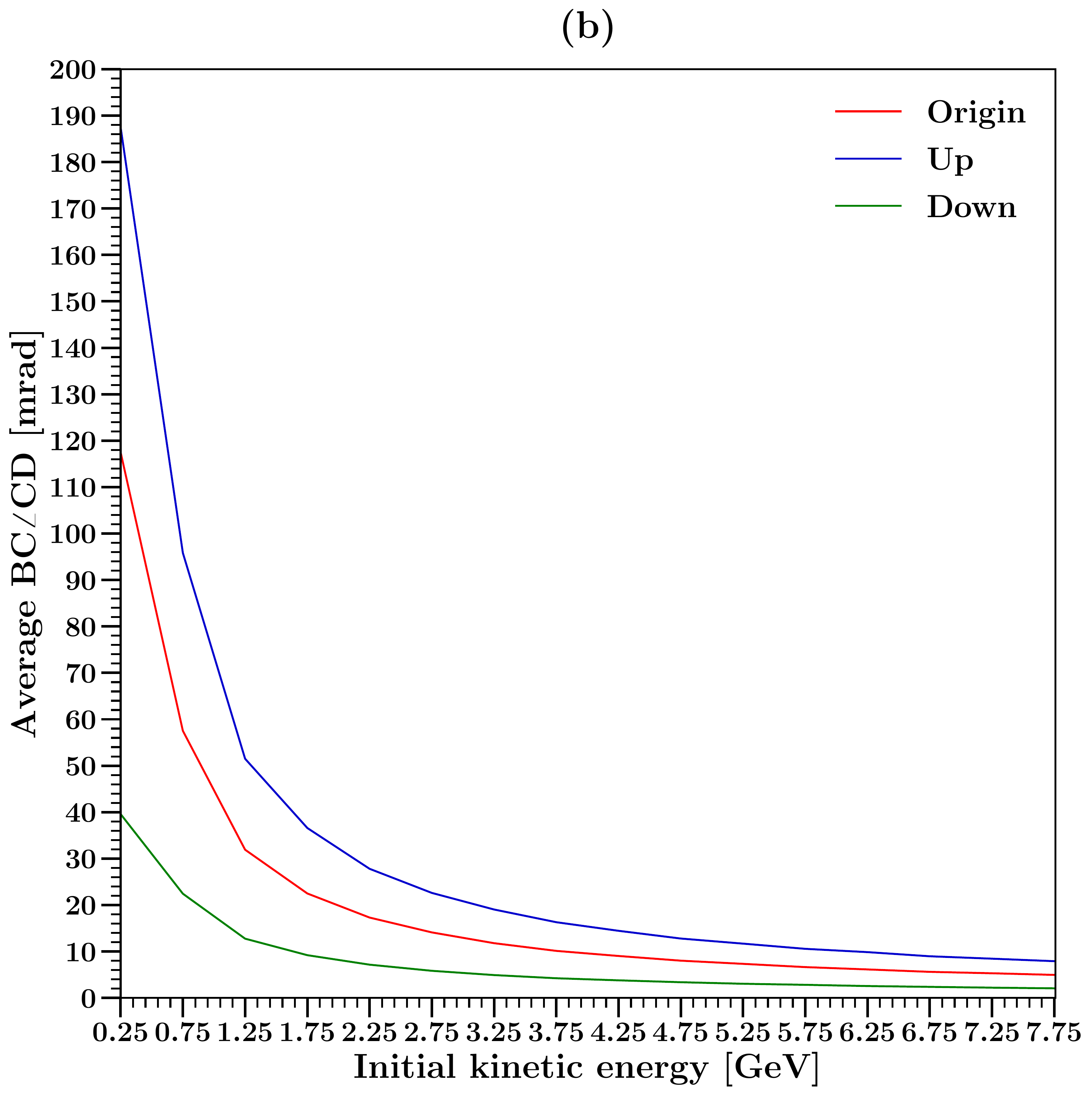}
\hskip 0.3cm
\includegraphics[width=7.34cm]{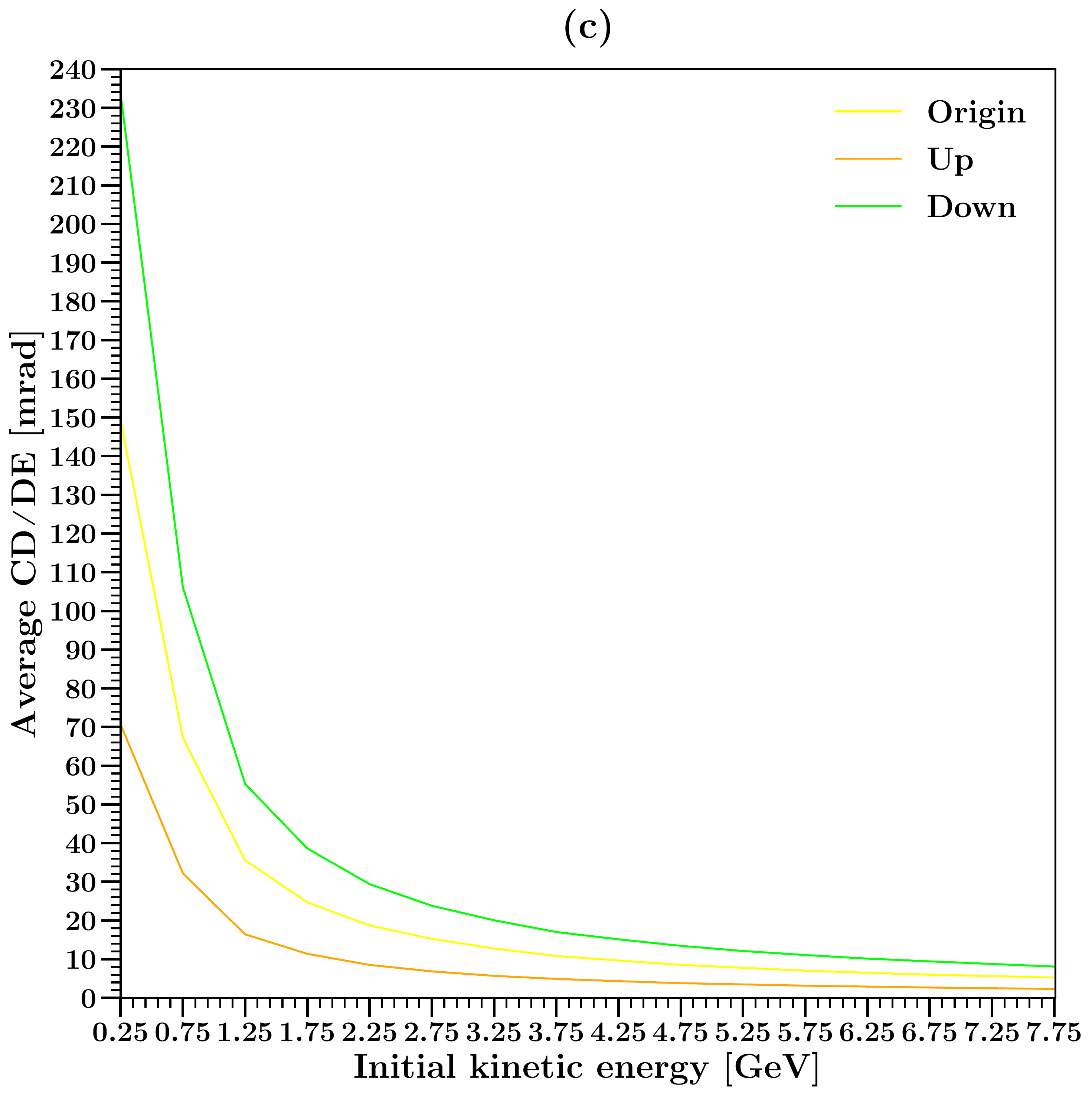}
\hskip 0.3cm
\includegraphics[width=7.34cm]{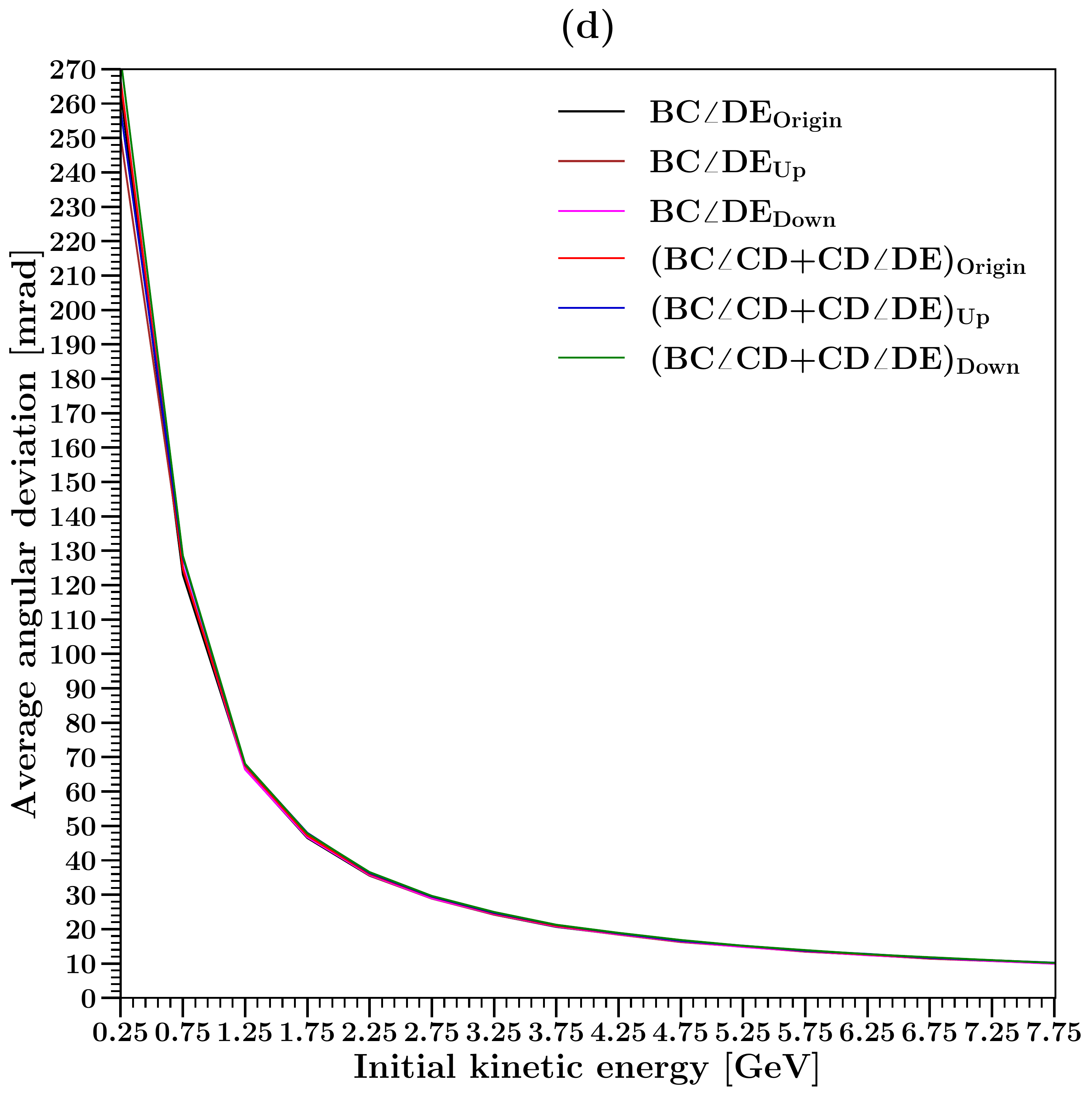}
\caption{Comparison between the average angular deviations over (a) $\rm  BC\angle DE$, (b) $\rm  BC\angle CD$, (c) $\rm  CD\angle DE$, and (d) sum of $\rm  BC\angle CD$ and $\rm  CD\angle DE$ for three different positions.}
\label{results}
\end{center}
\end{figure}

We commence our simulations with the scattering angle denoted by $\rm  BC\angle DE$ in order to investigate its position sensitivity versus the vertical displacement, and Fig.~\ref{results} (a) shows the average $\rm  BC\angle DE$ as a function of the kinetic energy. We observe that the average $\rm  BC\angle DE$ does not exhibit a tendency to vary with the vertical position change as demonstrated in Fig.~\ref{results} (a). 
\label{simulation outcomes}
Whereas the average $\rm BC\angle DE$ remains almost constant in spite of the spatial variation, the average interior non-adjacent angles indicated by $\rm  BC\angle CD$ and $\rm  CD\angle DE$ yield three distinct curves in the three different vertical positions as shown in Fig.~\ref{results} (b) and (c). Another reflection that we notice from Fig.~\ref{results} (b) and (c) is the opposite numerical trend among the opposite interior angles, which means that the average $\rm  BC\angle CD$ increases in terms of the vertical boost, while $\rm  CD\angle DE$ augments by the downward drop. At long last, we verify the triangular correlation as defined in Eq. (\ref{superpose}), and Fig.~\ref{results} (d) ratifies the equality between the scattering angle and the superposition of the interior non-adjacent angles through our GEANT4 simulations over three different positions.
\section{Conclusion}
In this study, we explore the triangular correlation of angular deviation by means of our GEANT4 simulations. Upon our simulation outcomes, we explicitly observe that the conventional scattering angle remains constant towards the position change of the target material, whereas the opposite interior angles exhibit differences due to this spatial variation, therewith hinting a beneficial property to be utilized for the image reconstruction purposes.
\label{conclusion}
\bibliographystyle{elsarticle-num}
\bibliography{triangular.bib}
\end{document}